\documentstyle[preprint,eqsecnum,aps]{revtex}
\tightenlines

\def\ga{\gamma}
\def\de{\delta}
\def\ep{\epsilon}

\def\De{\Delta}

\def\fr#1#2{{{#1} \over {#2}}}

\def\bra#1{\langle{#1}|}
\def\ket#1{|{#1}\rangle}

\def\frac#1#2{{\textstyle{{#1}\over {#2}}}}

\def\lsim{\mathrel{\rlap{\lower4pt\hbox{\hskip1pt$\sim$}}
    \raise1pt\hbox{$<$}}}
\def\gsim{\mathrel{\rlap{\lower4pt\hbox{\hskip1pt$\sim$}}
    \raise1pt\hbox{$>$}}}
\def\sqr#1#2{{\vcenter{\vbox{\hrule height.#2pt
         \hbox{\vrule width.#2pt height#1pt \kern#1pt
         \vrule width.#2pt}
         \hrule height.#2pt}}}}

\def\Re{\hbox{Re}\,}
\def\Im{\hbox{Im}\,}

\newcommand{\beq}{\begin{equation}}
\newcommand{\eeq}{\end{equation}}
\newcommand{\bea}{\begin{eqnarray}}
\newcommand{\eea}{\end{eqnarray}}
\newcommand{\rf}[1]{(\ref{#1})}

\begin{document}
\draft \preprint{}
\title{Theoretical correction to the neutral $B^0$ meson asymmetry\\ }
\author{J. C. Yoon \footnote{Email: jooyoon@indiana.edu} \\}
\address{
Physics Department\\
Indiana University\\
Bloomington, IN 47405, U.S.A.\\
}
\date{\today}
\maketitle
\begin{abstract}
Certain types of asymmetries in neutral meson physics have not
been treated properly, ignoring the difference of normalization
factors with an assumption of the equality of total decay width.
Since the corrected asymmetries in $B^0$ meson are different from
known asymmetries by a shift in the first order of CP- and
CPT-violation parameters, experimental data should be analyzed
with the consideration of this effect as in $K^0$ meson physics.
\end{abstract}
\section{Introduction}
To search for CP, CPT violation in neutral meson physics,
asymmetries are suggested in $K^0$ meson\cite{LeeWu} and heavy
meson\cite{CSanda,BSanda,DRosner,DW,D}. Though asymmetries of
decay modes from the same initial particles are not affected by
the normalization of decay modes, asymmetries of $P^0$(eg. $K^0$,
$B^0$, $D^0$) and $\overline P^0$ should be considered with the
difference of the normalization factors as in $K^0$ meson physics.
The normalization procedure will be derived in detail and the
assumption of the equality of total decay width will be
criticized. The correct theoretical asymmetries in $B^0$ meson to
be analyzed with experimental data will be suggested and
discussed.

\section{The normalization of neutral meson}
The effective Hamiltonian for the $P^0$-$\overline{P^0}$ system
has eigenvectors given by: \bea \ket{P_S} & = & [(1 + \ep_P +
\de_P)\ket{P^0} +(1 - \ep_P - \de_P)\ket{\overline{P^0}}]/{\sqrt
2}
\quad  \nonumber\\
\ket{P_L} & = & [(1 + \ep_P - \de_P)\ket{P^0} -(1 - \ep_P +
\de_P)\ket{\overline{P^0}}]/{\sqrt 2} \quad  \label{ii1} \eea The
parameter $\ep_P$ represents a CP violation with indirect T
violation, while the parameter $\de_P$ represents a CP violation
with indirect CPT violation.

We consider a coherent mixture of $P_S$ and $P_L$ whose amplitude
at its proper time $\tau$ is described by the wave function \bea
\Psi(\tau) & = & a_S e^{-i(m_S - i \fr {\gamma_S} {2}) \tau}
\ket{P_S} + a_L e^{-i(m_L - i \fr {\gamma_L} {2})\tau } \ket{P_L}
\eea The time evolution of the state is determined by an equation
of the Schr$\ddot{\textmd{o}}$dinger form: \bea i \fr d {d\tau}
\Psi & = & (M - \fr 1 2 i \Gamma)\Psi \eea where two Hermitian
operators M and $\Gamma$ called the mass and decay matrices. At
any instant decays will occur to specified final state $f$ with a
probability proportional to the square of the transition matrix
element. The total decay rate is given by summing it over all
final states $f$ consistent with energy and momentum conservation.
This must be compensated by a decrease in the probability of the
initial state\cite{Bell}\cite{Dalitz}. \bea - \fr d {d\tau}
[\Psi^\dagger \Psi] & = & \sum |\bra{f}T\ket{\Psi(\tau)}|^2
\label{ii6}\eea where $f$ is the final state and $\overline f$ is
the CP-conjugate state of $f$. \\
This equation should be normalized in a way that if we integrate
over proper time, we should get the number of initial particles.
To make Eq.\ \rf{ii6}, we can introduce the normalization factors.
\bea - \fr {1} {N} \fr d {d\tau} [\Psi^\dagger \Psi] & = & \fr
{\sum |\bra{f}T\ket{\Psi(\tau)}|^2} {\int \sum
|\bra{f}T\ket{\Psi(\tau)}|^2} \label{ii7}\eea where $N = - \int
\fr d {d\tau} [\Psi^\dagger \Psi] d\tau$.
The normalization factors is mentioned in Refs.\cite{PDG1} and
implemented in the monte carlo simulation of OPAL collaboration
\cite{OPAL}.

Let's consider two independent decays where the initial states are
$P^0$ and $\overline{P}^0$. Since these two decays are
independent, they should be normalized separately. \bea - \fr {1}
{N} \fr d {d\tau} [\Psi^\dagger \Psi] & = & \fr {\sum P_f(\tau) +
P_{\overline f}(\tau)} {\sum R_f + R_{\overline f}} \quad  \qquad
\nonumber\\
- \fr {1} {\overline{N}} \fr d {d\tau} [\overline{\Psi}^\dagger
\overline{\Psi}] & = & \fr {\sum \overline{P}_f(\tau) +
\overline{P}_{\overline f}(\tau)} {\sum \overline{R}_f +
\overline{R}_{\overline f}} \eea where $P_f =
|\bra{f}T\ket{\Psi(\tau)}|^2$, $R_{f} = \int d\tau P_{f}(\tau)$.
Note that we do not have to assume that the normalization factors
are the same for the independent decay modes of $P^0$ and
$\overline{P}^0$. The decay rate is not $P_f(\tau)$ but $\fr
{P_f(\tau)} {\sum R_f + R_{\overline f}}$. We still can compare
$P_f(\tau)$ and $P_{\overline f}(\tau)$, since they have the same
normalization factors. However we can not compare $P_f(\tau)$ and
$\overline{P}_f(\tau)$ without considering the normalization
factors. Since these normalization factors have CP- and
CPT-violation parameters with the opposite signs, we will have the
shift in the first order of CP- and CPT-violation parameters with
certain types of asymmetries. It is suggested that the total decay
widths of $P^0$ and $\overline{P}^0$ are equal by the CPT
theorem\cite{Dunietz} so that we do not have the difference in
normalization factors.
 \beq  \int_{0}^{\infty} d\tau \sum_f P_f(\tau) + P_{\overline f}(\tau)
 =
 \int_{0}^{\infty} d\tau \sum_f \overline{P}_f(\tau) +
\overline{P}_{\overline f}(\tau) \eeq  Note that the total decay
widths are the normalization factors. It is claimed that this
assumption is the only constraint, that is, we do not assume the
equality of decay width in each corresponding decay mode of $P^0$
and $\overline{P}^0$. Even though we accept the equality of total
decay width, we have to consider the difference of decay width in
a specific decay modes. However, the lifetime equality of CPT
symmetry was proved without consideration of the CP violation in
the mixing of neutral meson and tested where the mixing is not
involved\cite{Meyer}. With the mixing of neutral meson, we can not
clearly define the lifetime of neutral meson since it decays in
three different decay modes of short decay, long decay and
interference of these two. The total decay widths with the mixing
are also not as simple as in direct CPT symmetry due to indirect
CP- and CPT-violation. Since we still have the difference between
the total decay widths even without CPT violation as long as CP
symmetry is violated, CPT theorem does not guarantee the equality
of total decay widths in neutral meson with the mixing. The
equality of total decay widths is an assumption not based on CPT
theorem. Since we do not have any reason to assume the
normalization factors of two independent decay modes are the same,
I suggest to investigate asymmetries without assumption of the
equality of total decay widths.

We can build asymmetries of $P^0$ and $\overline{P}^0$ with three
different types of normalization methods. Experimentally, the
decay rate we get from the decay channel is $N \fr {P_f} {\sum R_f
+ R_{\overline f}}$ where $N$ is the number of initial particles
that we could get from the tagging. We can normalize by total
number of initial particles $N$ or by the number of events of its
own channel $N \fr {R_f + R_{\overline f}} {\sum R_f +
R_{\overline f}}$ or by those of other channel $N \fr {R_g +
R_{\overline g}} {\sum R_f + R_{\overline f}}$. \bea A_1 (\tau) &
\equiv & \fr { { \fr {\overline{P}_{\overline f}} {\sum
\overline{R}_{\overline f}+\overline{R}_f} -{\fr {P_f}{\sum
R_f+R_{\overline f}} }}} {{\fr {\overline{P}_{\overline f}} {\sum
\overline{R}_{\overline f}+\overline{R}_f}+{ \fr {P_f}{\sum
R_f+R_{\overline f}}}}} \eea

\bea A_2 (\tau) & \equiv & \fr {   { \fr {\overline{P}_{\overline
f}} {\overline{R}_{\overline f}+\overline{R}_f} -{\fr {P_f}{
R_f+R_{\overline f}} }}} {{\fr {\overline{P}_{\overline f}}
{\overline{R}_{\overline f}+\overline{R}_f}+{ \fr {P_f}{
R_f+R_{\overline f}}}}} \eea

\bea A_3 (\tau) & \equiv & \fr {   { \fr {\overline{P}_{\overline
f}} {\overline{R}_{\overline g}+\overline{R}_g} -{\fr {P_f}{
R_g+R_{\overline g}} }}} {{\fr {\overline{P}_{\overline f}}
{\overline{R}_{\overline g}+\overline{R}_g}+{ \fr {P_f}{
R_g+R_{\overline g}}}}} \eea Time-integrated asymmetries are
similar to these. Though the first type of normalization could be
attained in the experiment by production tagging, the effects due
to the different normalization factors are not clear unless all
decay modes are similar, since the difference of the normalization
factors depends on the different types of decay channels. The
third type of normalization was suggested in Refs.\cite{Dalitz}
which is normalized by $(K,\overline{K})^0 \rightarrow \pi\pi$
decay channel. CP and CPT test in CPLEAR was analyzed based on
this types of normalization, though the detailed calculations does
not agree with Refs.\cite{Dalitz}.
\section{B meson asymmetry}
Considering the decays to flavor-specific final states $f$ without
direct CP, CPT violation such as semileptonic decay mode \bea
\bra{f}T\ket{B^0} = F_f ~~~~,~~ & \bra{f}T\ket{\overline{B^0}} = 0
\quad  \nonumber \\
\bra {\overline f}T\ket {\overline{B^0}} = F_f^*~~~~,~~ & \bra
{\overline f}T\ket{B^0} = 0 \quad  \label{iid} \eea Due to mixing,
a produced $B^0$ can decay to the final state $\overline{f}$
(mixed event) in addition to the final state $f$ (unmixed event).
Here I followed the notation in \cite{Kostelecky}. The
time-dependent decay rates are given by: \bea P_f(\tau) & \equiv &
|\bra{f}T\ket{B(\tau)}|^2
\nonumber\\
&=& \frac 1 4 |F_f|^2 \{ (1+4\Re\de_B )\exp(-\ga_S \tau)
+(1-4\Re\de_B)\exp(-\ga_L \tau)
\nonumber\\
&&\quad\qquad + 2 \bigl[\cos \De m \tau - 4 \Im \de_B \sin \De m
\tau
      \bigr] \exp(-\ga \tau/2)
\}
\quad , \nonumber\\
\overline{P}_{\overline f} (\tau) & \equiv & |\bra{\overline
f}T\ket{\overline B(\tau)}|^2
\nonumber\\
&=& P_f(\de_B \rightarrow -\de_B)
\quad , \nonumber\\
P_{\overline f} (\tau) & \equiv & |\bra{\overline
f}T\ket{B(\tau)}|^2
\nonumber\\
&=& \frac 1 4 |F_f|^2 (1 - 4\Re\ep_B) \{ \exp(-\ga_S \tau) +
\exp(-\ga_L \tau) - 2\cos \De m \tau \exp(-\ga \tau /2) \}
\quad , \nonumber\\
\overline{P}_f(\tau) & \equiv & |\bra{f}T\ket{\overline
B(\tau)}|^2
\nonumber\\
&=& P_{\overline f} (\ep_B \rightarrow -\ep_B) \quad \label{iif}
\eea with $\ga = \ga_S + \ga_L$, $\De m = m_L - m_S$, $\De \ga =
\ga_S - \ga_L$, $ b^2 = (\Delta m)^2 + (\gamma^2)/ 4 $. The
time-integrated decay rates of $B^0$ and $\overline{B}^0$ are:
\bea R_f & \equiv & \int_{0}^{\infty} d\tau P_f(\tau)
\nonumber \\
& = & \frac 1 4 |F_f|^2 \bigg( \ga \Big[ \fr{1}{\ga_S \ga_L}
+\fr{1}{b^2} \Big] - \fr{\De \ga}{\ga_S \ga_L} 4\Re \de_B - \fr {
\De m}{b^2} 8\Im \de_B \bigg)
\quad , \nonumber\\
\overline{R}_{\overline f} & \equiv & \int_{0}^{\infty} d\tau
\overline{P}_{\overline f}(\tau) = R_f(\de_B \rightarrow -\de_B)
\quad , \nonumber\\
R_{\overline f} & \equiv & \int_{0}^{\infty} d\tau P_{\overline
f}(\tau)
\nonumber \\
& = & \frac 1 4 |F_f|^2 \ga \Big[ \fr{1}{\ga_S \ga_L} -\fr{1}{b^2}
\Big] (1 - 4\Re \ep_B)
\quad , \nonumber\\
\overline{R}_{f} & \equiv & \int_{0}^{\infty} d\tau
\overline{P}_f(\tau) = R_{\overline f}(\ep_B \rightarrow -\ep_B)
\quad . \label{iiib} \eea The normalization factors of $B^0$,
$\overline{B}^0$ are: \bea R_f+R_{\overline f}
 \equiv
\fr 1 4  |F_f|^2 (\Box - \triangle)
\nonumber \\
\overline{R} _{\overline f}+ \overline{R}_{f} \equiv  \fr 1 4
|F_f|^2 (\Box + \triangle) \eea where $\Box = \fr {2 \ga}{\ga_S
\ga_L}$, $\triangle = 4 \Re \ep_B \ga [ \fr 1 {\ga_S \ga_L} - \fr
1 {b^2}] + \fr {\De \ga} {\ga_S \ga_L} 4 \Re \de_B + \fr {\De
m}{b^2}8\Im \de_B$. The time-dependent asymmetries normalized by a
specific decay channel should be \beq A_{CP(T)}^{'}(\tau)
 = A_{CP(T)}(\tau) - \fr
{\triangle} {\Box} \eeq where the unnormalized time-dependent
asymmetries are: \bea A_{CP}(\tau) =  \fr {\overline{P}_f -
P_{\overline f}}{\overline{P}_f + P_{\overline f}}  & \simeq &
4\Re \ep_{B} \eea \bea A_{CPT}(\tau) = \fr
{\overline{P}_{\overline f} - P_{f}}{\overline{P}_{\overline f} +
P_{f}}& \simeq & \fr {4\Im \de_{B} \sin \De m\tau} { 1 + \cos \De
m \tau} \eea in the case of $\De \ga \simeq 0$. Note that these
unnormalized asymmetries are not attainable from experiment due to
the difference in normalization factors. Even we accept the
equality of total decay width, if it is normalized by a specific
decay mode, we get experimental data corresponding to
$A_{CP(T)}^{'}$ not $A_{CP(T)}$. In $B^{0}_d$ meson, the shift
$\fr {\triangle} {\Box}$ is $ \Im \de_B + 0.7 \Re \ep_B $ where
$\overline{\tau}_B = 1.548$ps $\De m_d = 0.472
ps^{-1}$\cite{PDG2}. Since this shift is not negligible, it should
be considered in fitting procedure. Since $A_{CPT}^{'}$ is also
sensitive to CP parameter $\ep_B$ besides CPT parameter $\de_B$,
even though $A_{CPT}^{'}$ is zero consistent, it doesn't mean CPT
symmetry is conserved.
\section{Conclusion}
Since the suggested asymmetries without the assumption of the
equality of total decay width are different from the known
asymmetries by a shift of the first order of CP- and CPT-violation
parameters, this effect can not be ignored in the analysis of CP
and CPT tests in neutral B meson.

\end{document}